\documentclass{PoS}

\title{Role of the Wilson mass parameter in the overlap Dirac
  topological charge density}

\ShortTitle{Role of the Wilson mass parameter in the overlap Dirac
  topological charge density}

\author{\speaker{Peter J. Moran},$^a$ Derek B. Leinweber$^a$
        and J. B. Zhang$^{ab}$\\
\llap{$^a$}Special Research Center for the Subatomic Structure of
   Matter (CSSM) and\\Department of Physics, University of Adelaide
   5005, Australia\\
\llap{$^b$}ZIMP and Department of Physics, Zhejiang University,
   Hangzhou, 310027,\\People's Republic of China\\
E-mail: \email{peter.moran@adelaide.edu.au}, \email{dleinweb@physics.adelaide.edu.au}, \email{jzhang@physics.adelaide.edu.au}}

\abstract{The negative Wilson mass parameter is an input parameter to
  the overlap Dirac operator.  We examine the extent to which the
  topological charge density, revealed by the overlap definition,
  depends on the value of the negative Wilson mass.  A strong
  dependence is observed, which can be correlated with the topological
  charge density obtained from the gluonic definition, with a variable
  number of stout-link smearing sweeps.  The results indicate that the
  freedom typically associated with fat-link fermion actions, through
  the number of smearing sweeps, is also present in the overlap
  formalism, through the freedom in the Wilson mass parameter.}

\FullConference{The XXVII International Symposium on Lattice Field
  Theory - LAT2009\\ July 26-31 2009\\ Peking University, Beijing,
  China}

\begin{document}

\section{Introduction}

By simulating the theory of quantum chromodynamics on a
four-dimensional space-time lattice, one can directly probe the
topological structure of the quantum vacuum.  An integral component of
the continuum theory is the realization of an exact chiral symmetry in
the massless limit.  Ideally, Lattice QCD calculations should also
observe this symmetry.  Unfortunately, simple transcriptions of the
fermion action explicitly break this symmetry at the order of the
lattice spacing.

The famous Nielsen-Ninomiya ``no-go'' theorem~\cite{Nielsen:1981hk}
explains the difficulties with implementing chiral symmetry on the
lattice.  In the early 80's, Ginsparg and
Wilson~\cite{Ginsparg:1981bj} proposed that the smoothest way to break
chiral symmetry on the lattice was to obey the Ginsparg-Wilson
relation,
\begin{equation}
  D \gamma_5 + \gamma_5 D = a D R \gamma_5 D \,.
  \label{GW}
\end{equation}
A popular solution to Eq.~(\ref{GW}) is the overlap Dirac
operator~\cite{Narayanan:1994gw,Neuberger:1997fp},
\begin{equation}
  D = \frac{m}{a} \left( 1 + \frac{D_W(-m)}{\sqrt{ D_W^{\dagger}(-m)
      \,D_W(-m) }} \right) \,,
  \label{overlap}
\end{equation}
where the Wilson-Dirac operator $D_w$, with a negative Wilson mass
$-m$, is the standard choice of input kernel.  The overlap operator is
known to observe an exact chiral symmetry on the lattice, and
satisfies the Atiyah-Singer index theorem, where the total topological
charge $Q$ is equal to the difference of Dirac zeromodes with opposite
chirality.

The topological charge density can be extracted from the trace of the
overlap operator,
\begin{equation}
  q_{ov}(x) = - \mathrm{Tr} \left[ \gamma_5 \left( 1 - \frac{a}{2\,m}
    D \right) \right] \,,
  \label{overlapqx}
\end{equation}
which allows one to investigate QCD vacuum structure through studies
of the $\langle q_{ov}(x) q_{ov}(0) \rangle$
correlator~\cite{Horvath:2002yn,Horvath:2005cv} and the topological
charge density itself~\cite{Horvath:2003yj,Ilgenfritz:2007xu}.  An
ultraviolet cutoff can be introduced through the spectral
representation~\cite{Horvath:2002yn,Koma:2005sw},
\begin{equation}
  q_{\lambda_{cut}}(x) = - \sum_{|\lambda|<\lambda_{cut}} \left( 1 -
  \frac{\lambda}{2\,m} \right) \psi_{\lambda}^{\dagger}(x) \gamma_5
  \psi_{\lambda}(x) \,.
  \label{spectralqx}
\end{equation}

We begin these proceedings by highlighting recent research
results~\cite{Ilgenfritz:2008ia} comparing the UV filtered overlap
topological charge density to the long established gluonic definition,
\begin{equation}
  q_{sm}(x) = \frac{g^2}{16\pi^2} \mathrm{Tr}~\left(
  F_{\mu\nu}\tilde{F}_{\mu\nu}\right) \,,
  \label{gluonicqx}
\end{equation}
obtained after smearing the gauge field.  Following this, we consider
the extent to which the negative Wilson mass parameter used in the
Wilson-Dirac input kernel of Eq.~(\ref{overlap}) affects the overlap
topological charge density.

\section{Stout-link smearing and the overlap Dirac operator}

The gluonic definition of the topological charge density in
Eq.~(\ref{gluonicqx}) is only valid on ``smooth'' gauge fields.  Many
prescriptions exist for smoothing a gauge field, however
discretization errors can skew the results.  For studies of QCD vacuum
structure, it is important to choose a topologically stable smearing
algorithm, such as the over-improved stout-link smearing
algorithm~\cite{Moran:2008ra}.  This is a modification of the original
stout-link algorithm, in which the standard plaquette is replaced by a
combination of plaquettes and rectangles, tuned to preserve
topological objects in the vacuum whilst smearing.  We use the
parameters proposed in Ref.~\cite{Moran:2008ra} and refer the reader
to that publication for full details of the algorithm and its
efficiency.  The topological charge density is then extracted using
Eq.~(\ref{gluonicqx}) in combination with an
$\mathcal{O}(a^4)$-improved field strength tensor.

In comparisons between the gluonic and overlap topological charge
densities, the gluonic definition may not always provide an integer
topological charge $Q$.  For this reason, it is common to apply a
multiplicative renormalization factor $Z$, $q_{sm}(x) \rightarrow
Z\,q_{sm}(x)$, where $Z$ is chosen such that $Q_{sm} \equiv \sum_x
q_{sm}(x)$ equals $Q_{ov}$, which is always an exact integer.  The
renormalization factors are typically close to $1$.  The interested
reader is encouraged to consult Ref.~\cite{Ilgenfritz:2008ia} for the
values.

\begin{figure}
  \begin{center}
    \includegraphics[height=0.4\textwidth,angle=90]{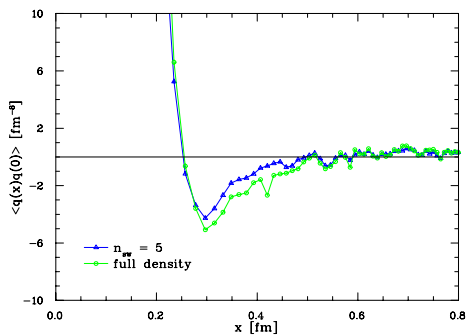}
  \end{center}
  \caption{The overlap $\langle q(x) q(0) \rangle$ correlator from the
    full topological charge density and the corresponding smeared
    correlator~\cite{Ilgenfritz:2008ia}.  The match can be further
    fine tuned by varying the smearing parameter.}
  \label{qxq0}
\end{figure}

We begin with a comparison of the $\langle q(x) q(0) \rangle$
correlator, which should be negative for any $x >
0$~\cite{Seiler:1987ig,Seiler:2001je,Horvath:2005cv}.  This was first
observed for the overlap in Ref.~\cite{Horvath:2005cv}, and it was
later shown in Ref.~\cite{Moran:2007nc}, that this behavior can be
reproduced with the gluonic definition.  Figure~\ref{qxq0} shows the
overlap correlator and the best smeared match using a smearing
parameter of $\rho = 0.06$ with 5 sweeps of over-improved stout-link
smearing.  We note that by varying the magnitude of $\rho$ this match
can be further fine tuned.  This matching of the $\langle q(x) q(0)
\rangle$ enables high statistics studies using the computationally
efficient smearing algorithm, and was used in Ref.~\cite{Moran:2008qd}
to explore the effect of dynamical quarks on QCD vacuum structure.

\begin{figure}
  \begin{center}
    \begin{tabular}{cc}
      \includegraphics[width=0.35\textwidth,angle=0]{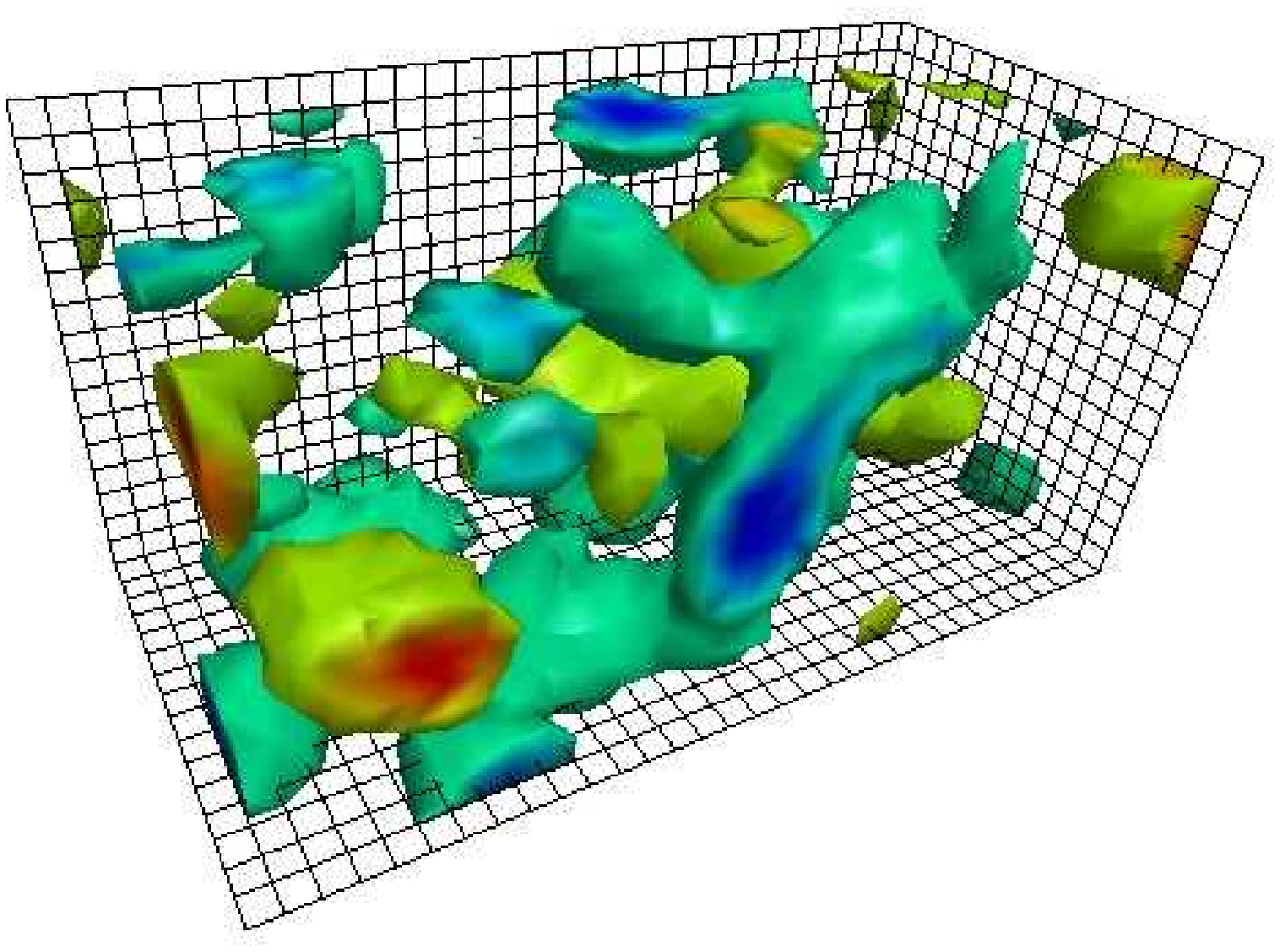} &
      \includegraphics[width=0.35\textwidth,angle=0]{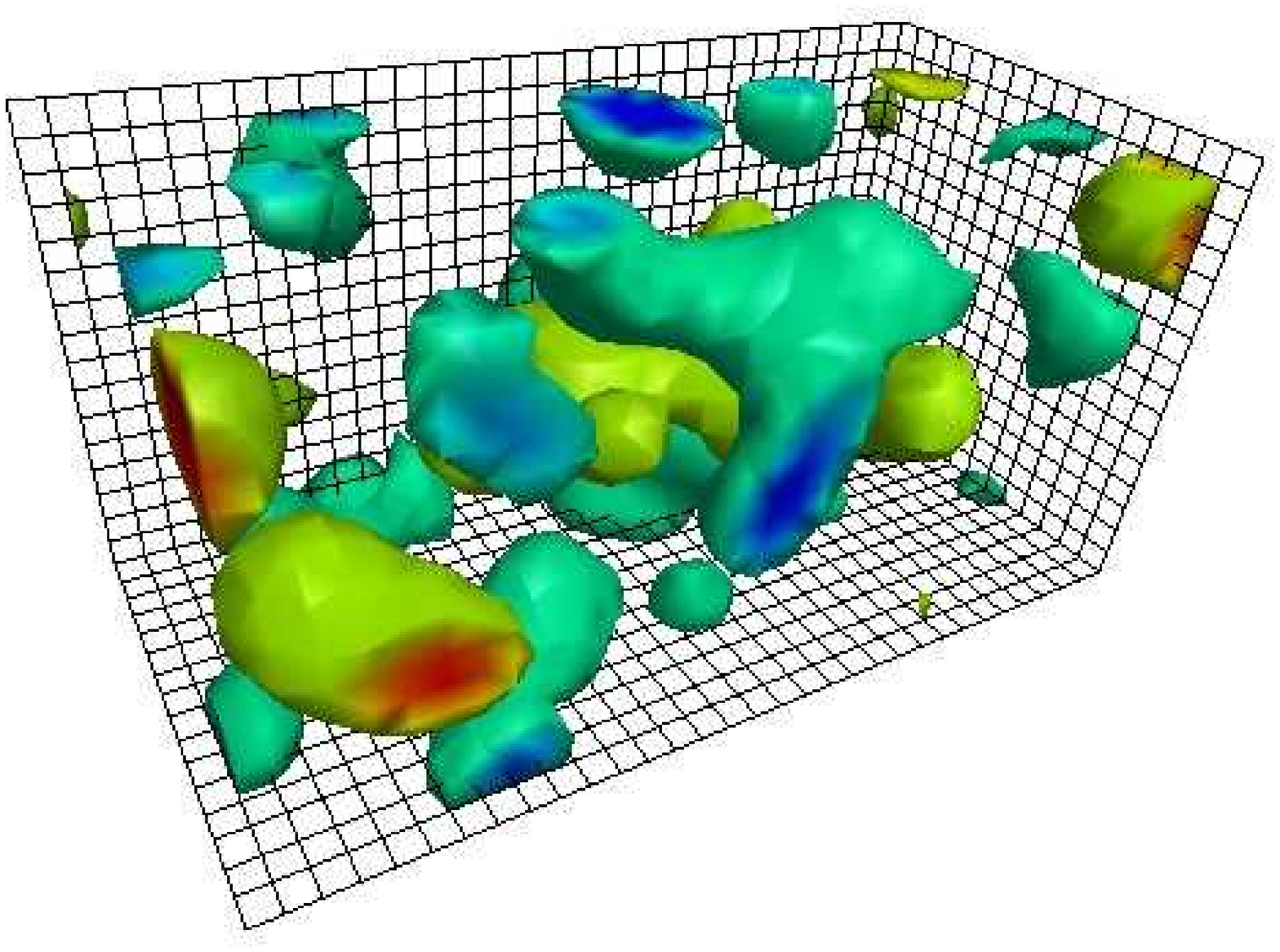}
    \end{tabular}
  \end{center}
  \caption{The overlap topological charge density (left) calculated
    using an UV cutoff of $\lambda_{\rm cut} = 634$~MeV.  On the right
    is the best smeared match found using $48$ sweeps of over-improved
    stout-link smearing~\cite{Ilgenfritz:2008ia,Moran:2008ra}.}
  \label{634MeV}
\end{figure}

Ilgenfritz \emph{et al.}~\cite{Ilgenfritz:2008ia} further explored the
correlation between the gluonic and overlap topological charge density
using the spectral representation of the overlap operator,
Eq.~(\ref{spectralqx}).  They found a direct connection between the
number of stout-link smearing sweeps applied to the gauge field and
the strength of the UV cutoff.  In Fig.~\ref{634MeV} we present a
sample of that work, the best match for a cutoff of $\lambda_{\rm cut}
= 634$~MeV.

\section{Role of the negative Wilson mass parameter}

The (negative) Wilson mass parameter enters the definition of the
overlap operator through the Wilson-Dirac input kernel.  At tree level
the allowed range for the Wilson mass is $0 < m < 2$.  However, when
working at a finite lattice spacing $a$, one must also have $m
\gtrsim 1.0$~\cite{Edwards:1998sh}.  Converting this to the more
standard $\kappa$ parameter gives an allowed range of $1/6 \lesssim
\kappa < 1/4$, since at tree level,
\begin{equation}
  \kappa \equiv \frac{1}{2\,(-m)\,a + 8\,r} \,.
\end{equation}

By varying $m$ within this range, one has access to a fermionic probe
of the gauge background at different scales $\sim
1/m$~\cite{Neuberger:1997fp}.  Previous studies have investigated how
the total topological charge and topological susceptibility depend on
the value of the Wilson
mass~\cite{Narayanan:1994gw,Edwards:1998sh,Narayanan:1997sa,DelDebbio:2003rn}.
We now extend this work to include an analysis of the topological
charge density itself.  This should provide some useful physical
insights since the low-lying modes of the Dirac operator are strongly
correlated with the topological charge
density~\cite{Ilgenfritz:2008ia,Kusterer:2001vk}.

The overlap Dirac operator is an expensive calculation so we consider
a single spatial slice from some representative $16^3 \times 32$
configurations.  Five $\kappa$ values, $0.17, 0.18, 0.19, 0.21$ and
$0.23$ are used to investigate the effect of the Wilson mass on
the topological charge density.  As before, we monitor the changes in
$q_{ov}(x)$ through direct visualizations.

\begin{figure}
  \begin{center}
  \begin{tabular}{c}
    \includegraphics[width=0.2\textwidth]{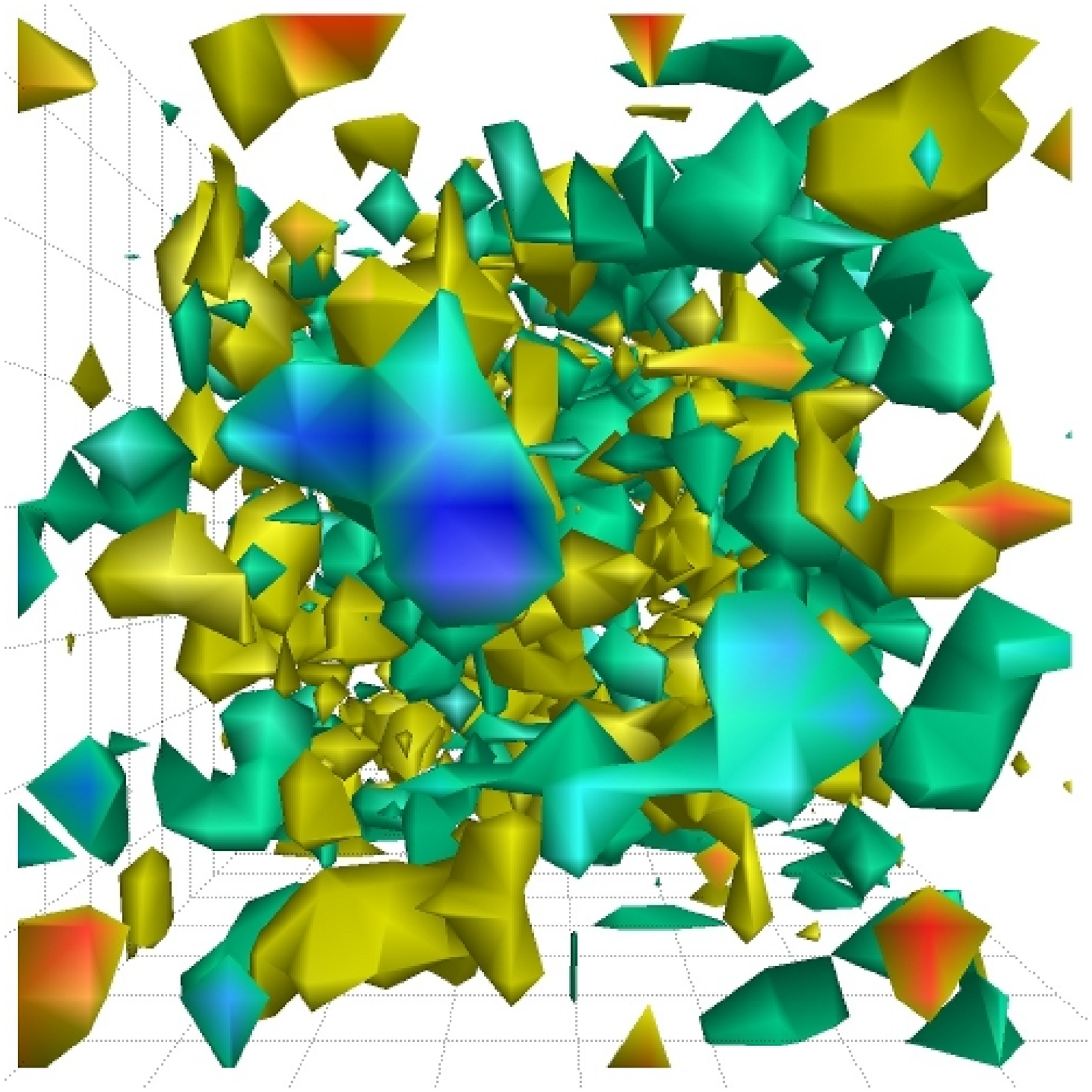} 
    \includegraphics[width=0.2\textwidth]{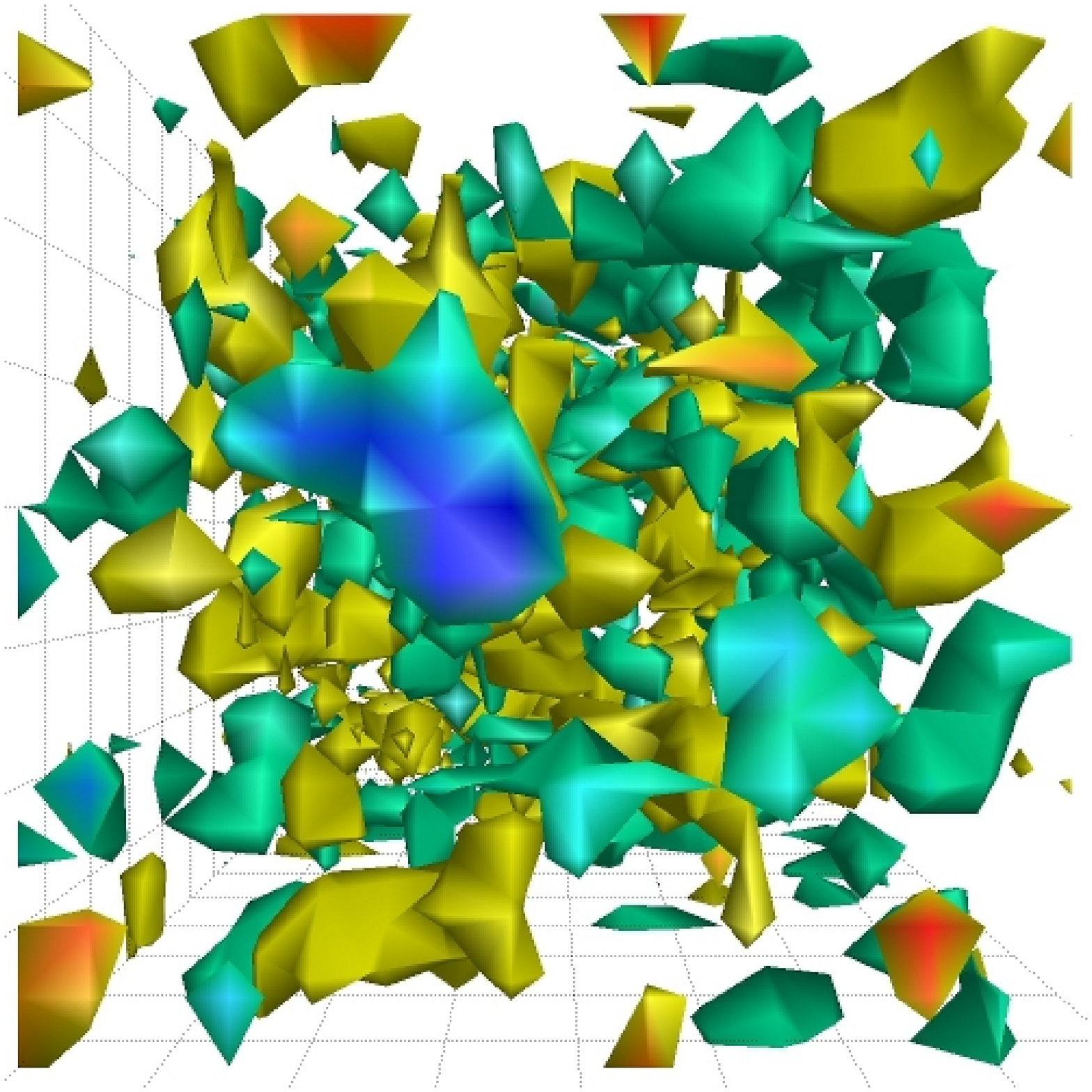} 
    \includegraphics[width=0.2\textwidth]{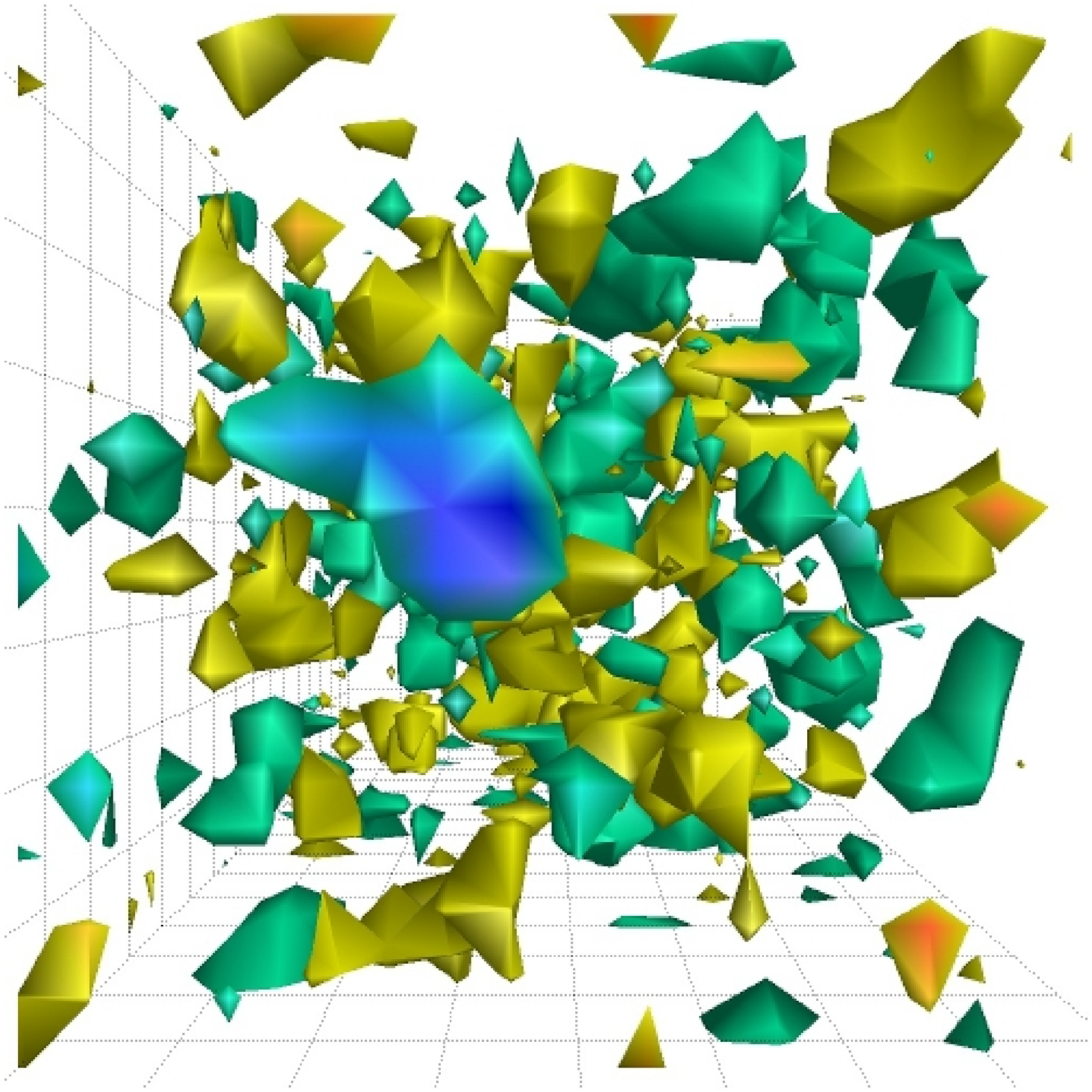} \\
    \includegraphics[width=0.2\textwidth]{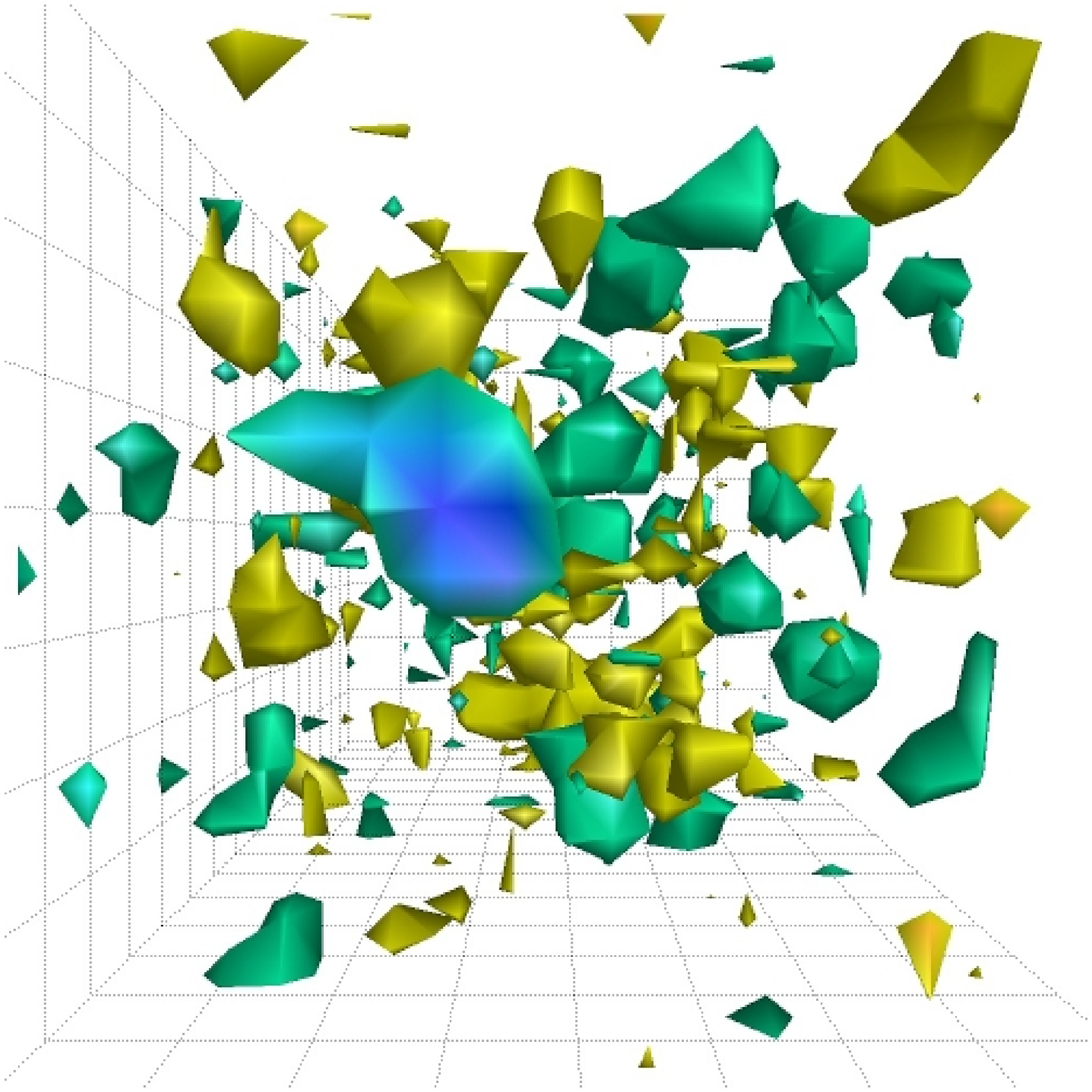}
    \includegraphics[width=0.2\textwidth]{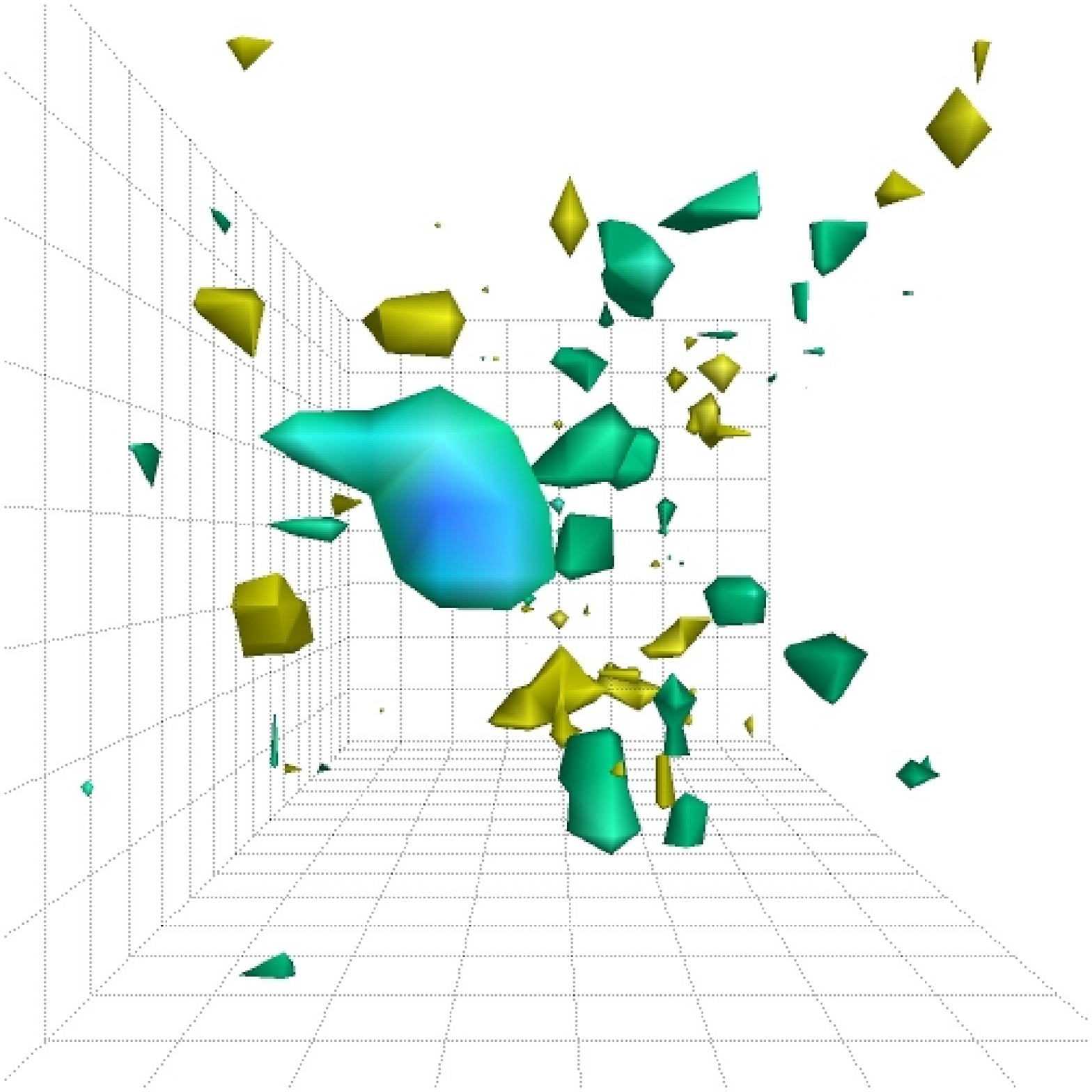} 
  \end{tabular}
  \end{center}
  \caption{The overlap topological charge density $q_{ov}(x)$
    calculated with five choices for the Wilson mass $m$.  From left
    to right, we have $\kappa = 0.23$, $0.21$, and $0.19$ on the first
    row, with $0.18$, and $0.17$ on the second.  There is a clear
    dependence on the value of $m$ used, with larger values revealing
    a greater amount of topological charge density.}
  \label{fig:overlapqx}
\end{figure}

The topological charge densities, for the five choices of $\kappa$,
are presented in Fig.~\ref{fig:overlapqx}.  A clear dependence on the
Wilson mass is apparent from the figures, with smaller values of $m$
revealing greater topological charge density.  This is consistent with
expectations, since as $m$ is increased the Dirac operator becomes
more sensitive to smaller topological objects.

The changes in $q_{ov}(x)$ as $m$ is varied appear very similar to the
way that the gluonic topological charge density depends on the number
of smearing sweeps applied, and we now quantify this connection.  We
use a relatively weak smearing parameter\footnote{To be compared with
  the usual $\rho = 0.06$ for over-improved stout-link smearing, or
  $\rho = 0.1$ for standard stout-link smearing} of $\rho = 0.01$, and
as before, applying a multiplicative renormalization factor.  We
consider two choices for $Z$, firstly a \emph{calculated} $Z$,
determined using,
\begin{equation}
  Z = \frac{ \sum_x |q_{ov}(x)| }{ \sum_x |q_{sm}(x)| } \,.
  \label{calcZ}
\end{equation}
This will be compared with a \emph{fitted} $Z$ where $Z$ is found by
minimizing,
\begin{equation}
  \sum_x |q_{ov}(x) - Z\,q_{sm}(x)| \,.
\end{equation}

\begin{figure}
  \begin{center}
  \begin{tabular}{cc}
    \includegraphics[width=0.2\textwidth]{c002_k23.eps} &
    \includegraphics[width=0.2\textwidth]{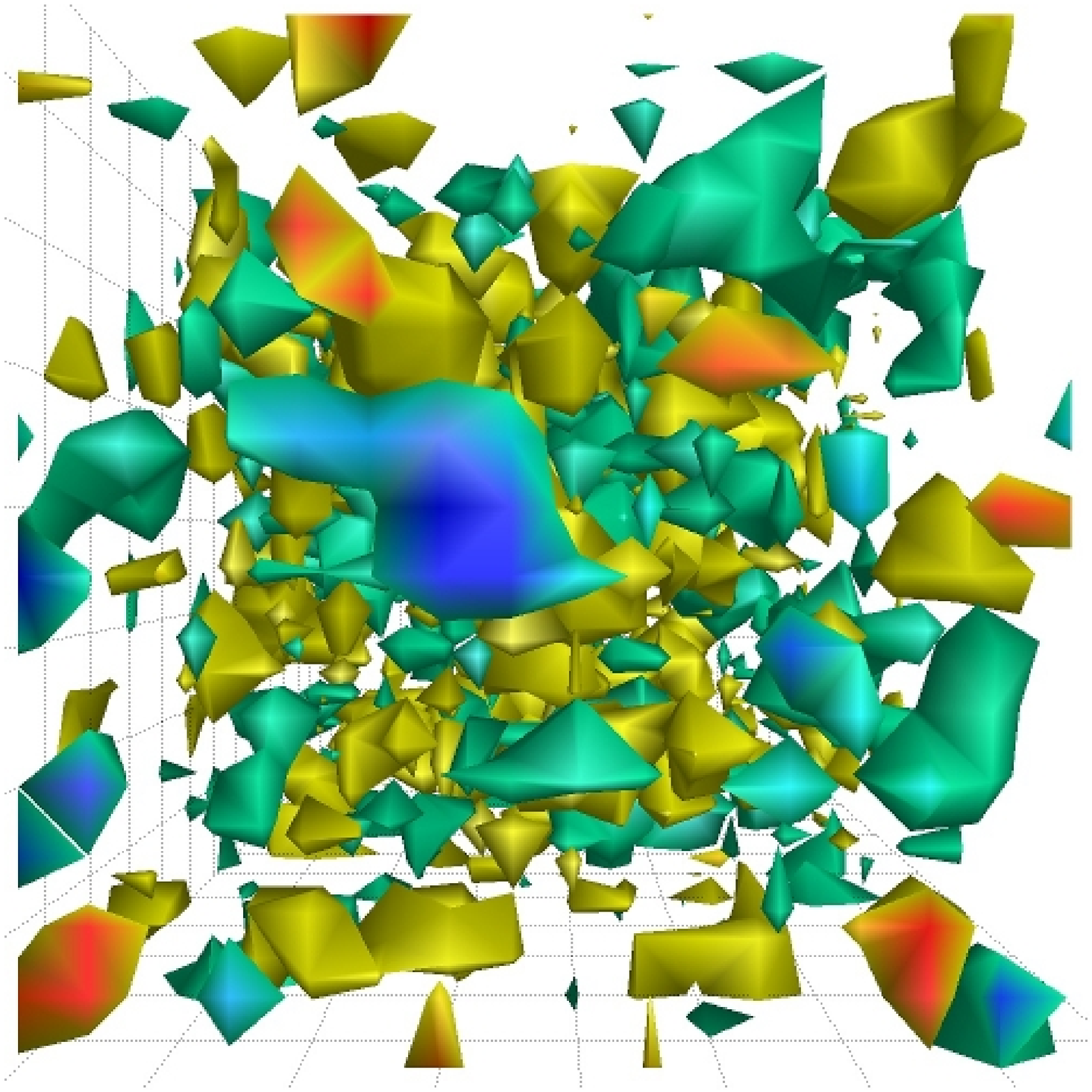} \\
    $\kappa = 0.23$ & $n_{sw} = 22$ \\
    \includegraphics[width=0.2\textwidth]{c002_k19.eps} &
    \includegraphics[width=0.2\textwidth]{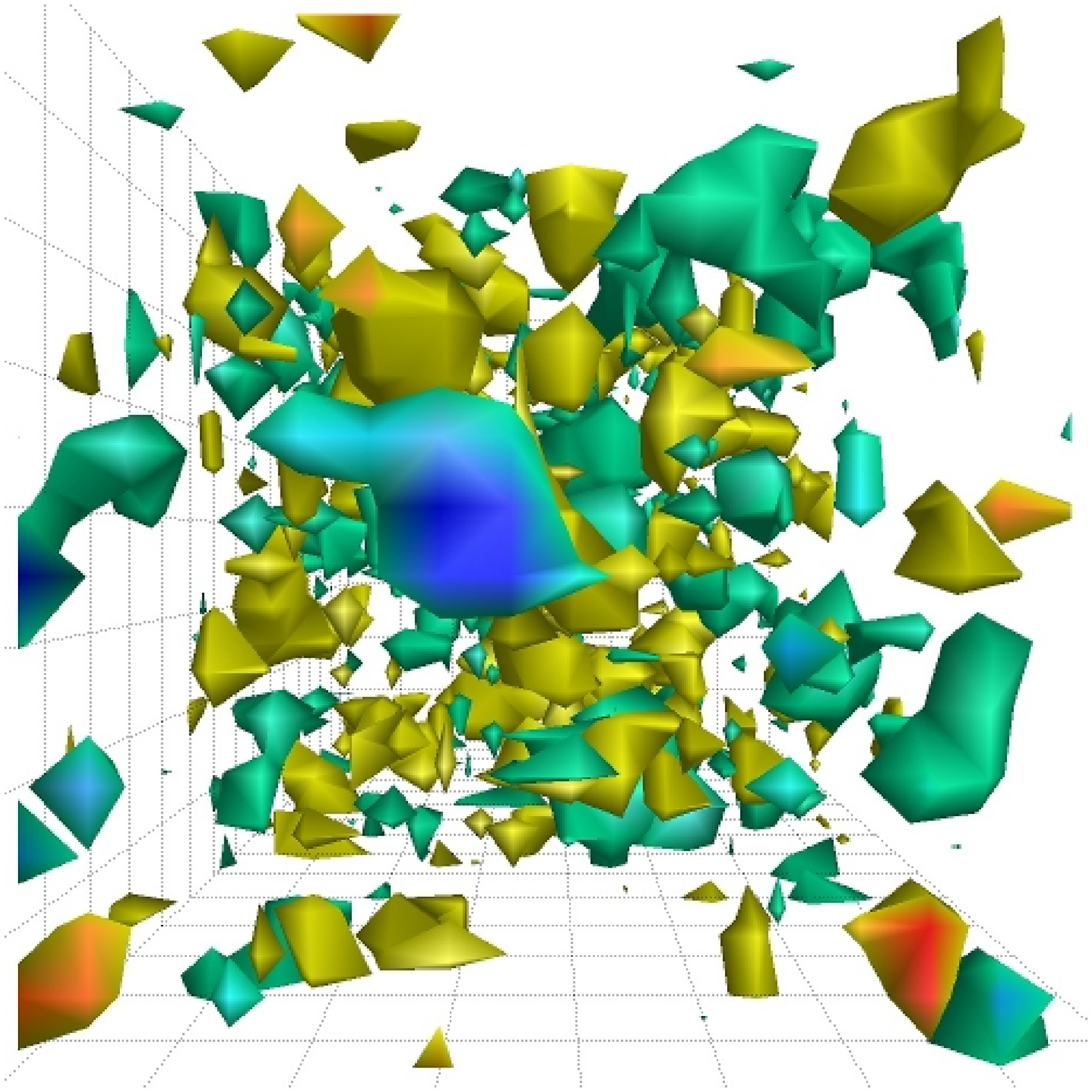} \\
    $\kappa = 0.19$ & $n_{sw} = 25$ \\
    \includegraphics[width=0.2\textwidth]{c002_k17.eps} &
    \includegraphics[width=0.2\textwidth]{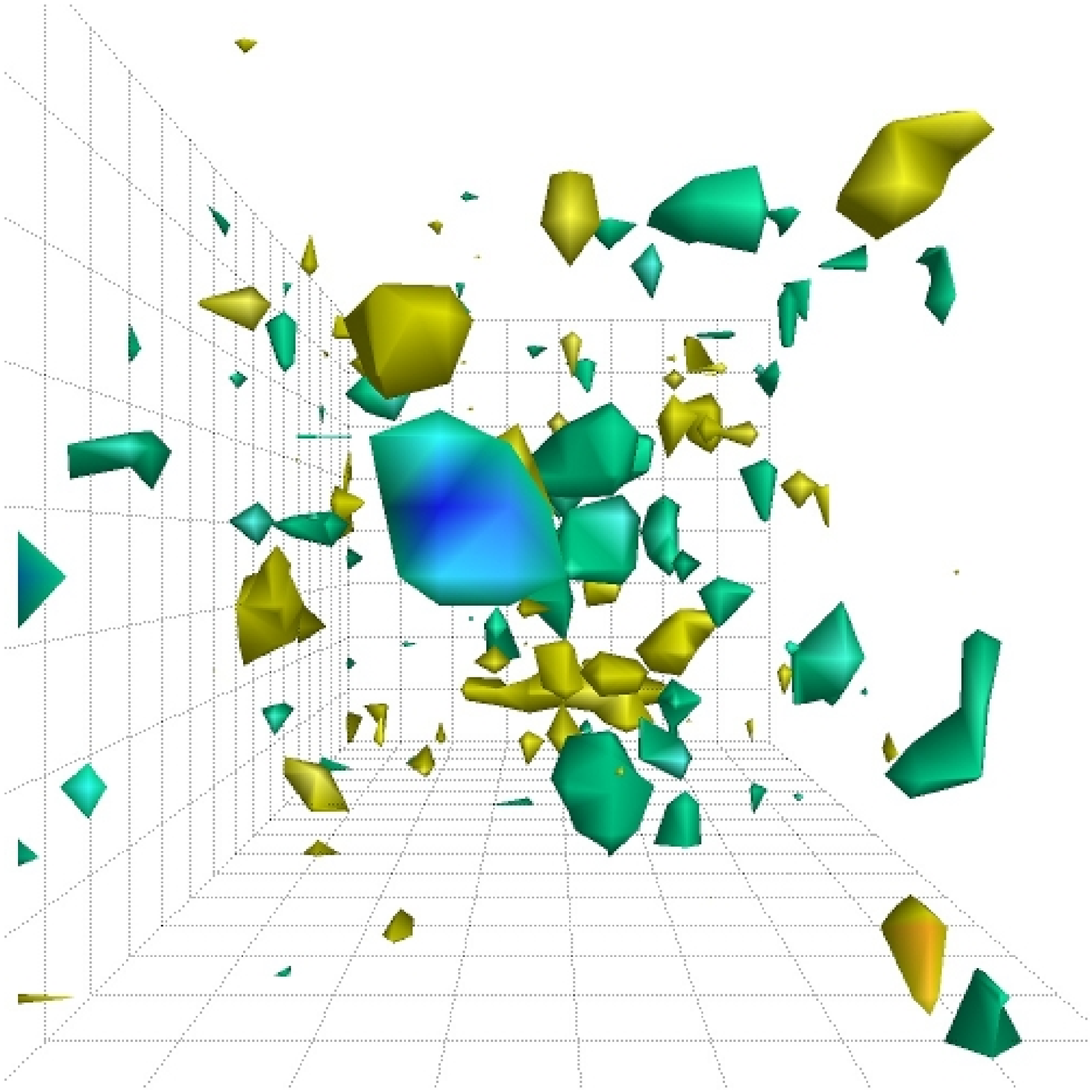} \\
    $\kappa = 0.17$ & $n_{sw} = 28$
  \end{tabular}
  \end{center}
  \caption{The best smeared matches (right) compared with three of the
    overlap topological charge densities (left) in order of increasing
    $m$, where $q_{sm}(x)$ is renormalized using the calculated $Z$.}
  \label{bestmatches}
\end{figure}

\begin{table}
  \begin{center}
    \begin{tabular}{l@{\hspace{2cm}}r}
      \begin{tabular}{ccc}
        $\kappa$ & $n_{sw}[\mathrm{calc.}\ Z]$ & $n_{sw}[\mathrm{fitted}\ Z]$ \\
        \hline
        0.17 & 28 & 29  \\
        0.18 & 26 & 27  \\
        0.19 & 25 & 25  \\
        0.21 & 23 & 23  \\
        0.23 & 22 & 23 
      \end{tabular}
      &
      \begin{tabular}{ccc}
        $\kappa$ & $n_{sw}[\mathrm{calc.}\ Z]$ & $n_{sw}[\mathrm{fitted}\ Z]$ \\
        \hline
        0.17 & 29 & 30  \\
        0.18 & 26 & 27  \\
        0.19 & 25 & 25  \\
        0.21 & 23 & 23  \\
        0.23 & 22 & 22 
      \end{tabular}
    \end{tabular}
  \end{center}
  \caption{The best smeared matches for all five $\kappa$ values
    considered.  Results for two configurations are reported in Table
    (a) (left) and (b) (right). There is a definite correlation
    between the choice of the Wilson mass and the number of smearing
    sweeps required to match the topological charge density.}
  \label{compareZ}
\end{table}

The overlap topological charge densities along with the corresponding
best matches found using the calculated $Z$ are shown in
Fig.~\ref{bestmatches}.  Due to a lack of space we show only the
largest, smallest and middle $\kappa$ values.  As the Wilson mass is
increased, and non-trivial topological charge fluctuations are
removed, a greater number of smearing sweeps are needed in order to
recreate the topological charge density.  A summary of the best
matches are provided in Table~\ref{compareZ}a along with the results
obtained using a fitted $Z$ value.  We note the good agreement between
the two choices for fixing the renormalization factor.  Results for a
second configuration are provided in Table~\ref{compareZ}b.  A
comparison reveals little difference in the number of smearing sweeps
required to match the overlap $q(x)$.

\section{Conclusion}

The overlap topological charge density displays a clear dependence on
the value of the Wilson mass used in the calculation.  Although this
was not unexpected, we are also able to show how one can correlate the
value of $\kappa$ used in the overlap to a specific number of
stout-link smearing sweeps.  This implies an intimate relationship
between the number of smearing sweeps and the value of $\kappa$ in the
overlap formalism.

We have shown how the smoothness of the gauge field as seen by the
overlap operator depends on the value of the Wilson mass.  This is
similar to fat-link fermion actions where the gauge links are smeared
with a small number of smearing sweeps and the smoothness depends on
the number of applied sweeps.  The results indicate that the freedom
typically associated with fat-link fermion actions, through the number
of smearing sweeps, is also present in the overlap formalism, through
the freedom in the Wilson mass parameter.  However the number of
smearing sweeps relevant to the overlap operator is small.  This
connection will be further explored in a forthcoming
publication~\cite{Moran:Future}.

\acknowledgments

We thank both eResearch SA and the NCI National Facility for generous
grants of supercomputer time which have enabled this project.  This
work is supported by the Australian Research Council. J. B. Zhang is
partly supported by Chinese NSFC-Grant No.~10675101 and 10835002.

\providecommand{\href}[2]{#2}\begingroup\raggedright\endgroup


\end{document}